\documentclass[twocolumn]{el-author}

\usepackage{diagbox}

\newcommand{\ua}{\uparrow}
\newcommand{\nc}{\newcommand}
\nc{\da}{\downarrow} \nc{\hc}{\hat{c}} \nc{\hS}{\hat{S}}
\nc{\bra}{\langle} \nc{\ket}{\rangle} \nc{\eq}{equation (\ref}
\nc{\h}{\hat} \nc{\hT}{\h{T}}\nc{\be}{\begin{eqnarray}}
\nc{\ee}{\end{eqnarray}}\nc{\rd}{\textrm{d}}\nc{\e}{eqnarray}\nc{\hR}{\hat{R}}\nc{\Tr}{\mathrm{Tr}}
\nc{\tS}{\tilde{S}}\nc{\tr}{\mathrm{tr}}\nc{\8}{\infty}\nc{\lgs}{\bra\ua,\phi|}\nc{\rgs}{|\ua,\phi\ket}
\nc{\hU}{\hat{U}}\nc{\lfs}{\bra\phi|}\nc{\rfs}{|\phi\ket}\nc{\hZ}{\hat{Z}}\nc{\hd}{\hat{d}}\nc{\mD}{\mathcal{D}}
\nc{\bd}{\bar{d}}\nc{\bc}{\bar{c}}\nc{\mc}{\mathcal}\nc{\ea}{eqnarray}\nc{\mG}{\mathcal{G}}\nc{\bce}{\begin{center}}
\nc{\ece}{\end{center}}
\date{10th July 2019}

\begin{document}

\title{Blind Packing Ratio Estimation for Faster than Nyquist Signaling based on Deep Learning}

\author{Peiyang Song, Fengkui Gong and Qiang Li}

\abstract{This letter proposes a blind packing ratio estimation for faster than Nyquist (FTN) signaling based on state-of-the-art deep learning (DL) technology. The packing ratio is a vital parameter to obtain the real symbol rate and recover the origin symbols from the received symbols by calculating the intersymbol interference (ISI). To the best of our knowledge, this is the first effective estimation approach for packing ratio in FTN signaling and has shown its fast convergence and robustness to signal-to-noise ratio (SNR) by numerical simulations. Benefiting from the proposed blind estimation, the packing-ratio-based adaptive FTN transmission without dedicate channel or control frame becomes available. Also, the secure FTN communications based on secret packing ratio can be easily cracked.}

\maketitle

\section{Introduction}
The last several decades have witnessed the exponential growth of wireless devices and data traffic. With the increasingly demanding requirement for spectral resources, a promising technology named FTN is rediscovered and has started its employment in mobile and satellite communications.

In conventional Nyquist-criterion systems, the symbol duration is always set as $T_N=1/(2W)$, where $W$ is the bandwidth. In the 1970s, Mazo \cite{1} firstly proved that by reducing the symbol duration to $T=\alpha T_N (0<\alpha<1)$, the FTN signaling can achieve a higher transmission rate than conventional Nyquist-criterion design without loss of bit error rate (BER) performance in additive white Gaussian noise (AWGN) channel. The scaling factor $\alpha$ is called the packing ratio. Until now, various signal detections for FTN \cite{2, 3} have been proposed. However, as far as we know, the blind packing ratio estimation for FTN signaling has not been studied yet.

In practical, both the signal downsampling and detection in the receiver require the accurate value of packing ratio $\alpha$ that is usually considered as a known condition in conventional researches. However, this may be not proper in some scenarios. For example, when an eavesdropper tries to eavesdrop an FTN-aided communication with unknown $\alpha$, the existing detection algorithms will not work due to the unknown real symbol rate and ISI. Also, when packing-ratio-based adaptive FTN transmission is considered, a dedicated channel or control frame must be employed to inform the receiver of the current $\alpha$. This may violate the original intention of FTN to improve spectrum utilization.

In recent years, a new trend has appeared to merge the two technologies of communications and DL. For example, \cite{4} improves decoding for linear codes with DL technologies. \cite{5} proves that deep learning is a promising tool for channel estimation and signal detection in wireless communications with complicated channel distortion and interference. Inspired by the challenges of packing ratio estimation and the increasingly wide application of DL in communications, we propose a DL-based blind packing ratio estimation for FTN signaling.

\section{System Model of FTN}

\begin{figure}[h]
	\centering{\includegraphics[width=80mm]{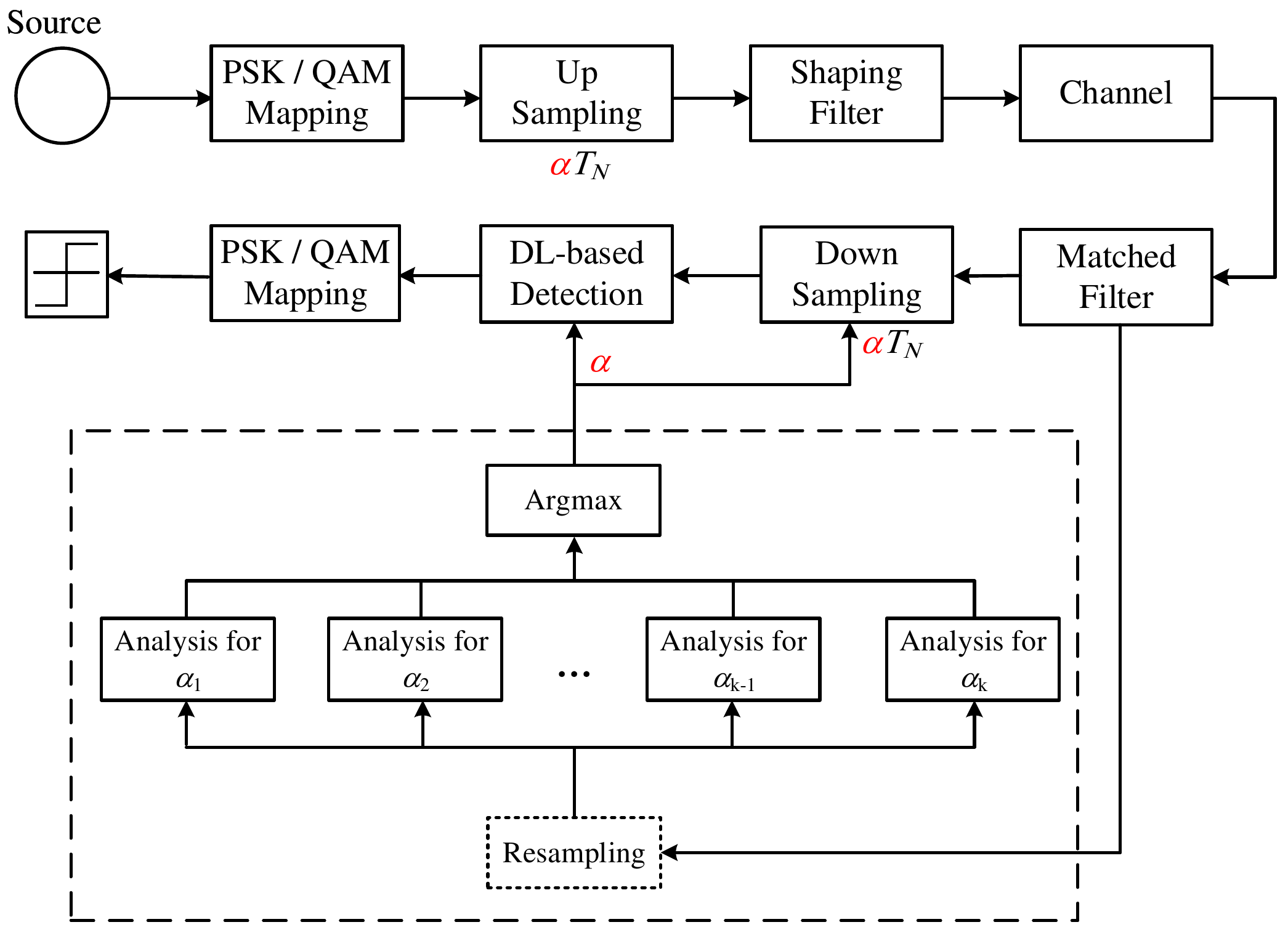}}
	\caption{Block diagram of the FTN signaling with the proposed DL-based blind packing ratio estimation
		\source{}}
	\label{fig:block_diagram}
\end{figure}

The block diagram of the FTN signaling with the proposed DL-based packing ratio estimation has been illustrated in Fig \ref{fig:block_diagram}. In this letter, we consider the communication system with the binary phase shift keying (BPSK) modulation scheme and AWGN channel. Here, we directly give the received symbols which have passed through the matched filter as

\begin{align} \label{eq:received_samples}
	y_{n}  & =\sqrt{E_{s}}\sum_{k=-\infty}^{+\infty}x_{k}g(n\alpha T_{N}-k\alpha T_{N})+\widetilde{n}(n\alpha T_N)\nonumber \\
	& =\underset{ISI\,from\,previous\,L-1\,symbols}{\underbrace{\sqrt{E_{s}}\sum_{k=-\infty}^{n-1}x_{k}g\left(\left(n-k\right)\alpha T_{N}\right)}}+\sqrt{E_{s}} x_{n}g(0)\nonumber \\
	& \quad+\underset{ISI\,from\,upcoming\,L-1\,symbols}{\underbrace{\sqrt{E_{s}}\sum_{k=n+1}^{+\infty}x_{k}g\left(\left(n-k\right)\alpha T_{N}\right)}}+\widetilde{n}\text{\ensuremath{\left(n\alpha T_N)\right)}},
\end{align}
where $g(t)=\int h(x)h(t-x)dx$, $\widetilde{n}(t)=\int n(x)h(t-x)dx$. $h(t)$ is the function of the shaping filter. $n(t)$ is a zero mean complex-valued Gaussian random process
with variance $\sigma^{2}$. And $E_{s}$ is the average energy of constellation symbols. As seen from (\ref{eq:received_samples}), $\alpha$ is an important parameter to operate the downsampling and calculate the ISI between different received symbols.

\section{The Proposed DL-based Blind Packing Ratio Estimation}
It is noteworthy that $T_N$ and the matched filter are considered as known parameters because they will be determined in advance or be obtained by the eavesdropper with spectrum analysis. For the convenience of simulation and explanation, in this letter, we consider square root raised cosine (SRRC) shaping filter with 20 times upsampling.

Due to the small enough difference among received symbols with different $\alpha$, employing deep neural network (DNN) \cite{6} directly to classify the $\alpha$ is not feasible, which has been proved by our experiments. Fig. \ref{fig:block_diagram} illustrates the architecture of our proposed DL-based blind packing ratio estimation. The analyses for different $\alpha_k$ values are employed independently, where the input and the output are respectively the symbols after the matched filter and the count of true decisions for whether $\alpha=\alpha_k$. Finally, the $\alpha_k$ with the largest count of true decisions during a certain time will be considered as the estimation result. It is worth noting that the output of the matched filter may be resampled to make the downsampling interval an integral multiple of $T_N/I$, where $I$ is the number of samples per symbol. For example, $I=20$ may be employed when the possible $\alpha$ values include 0.75 and 0.8.

\begin{figure}[h]
	\centering{\includegraphics[width=80mm]{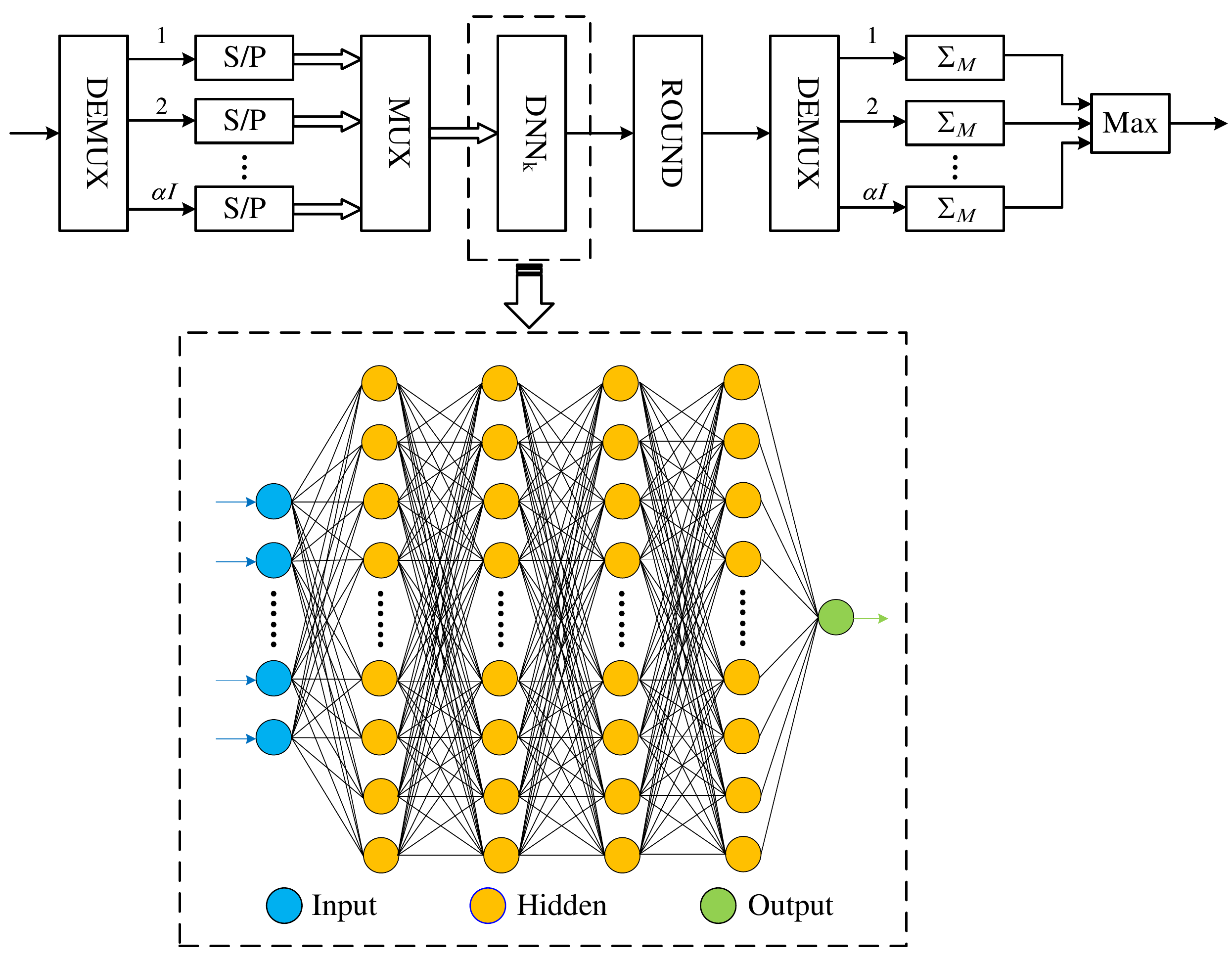}}
	\caption{The detailed structure of the analysis for $\alpha_k$ in the proposed blind estimation
		\source{}}
	\label{fig:subblock}
\end{figure}

The detailed structure of the analysis for $\alpha_k$ is demonstrated in Fig. \ref{fig:subblock}. $\alpha_k I$ branches are employed for the multiplexer (MUX) and the demultiplexer (DEMUX). Symbols downsampled by interval $\alpha_k T_N$ in different locations are input into the DNN$_k$ respectively. The output of the DNN$_k$, which can be regarded as the probability of $\alpha_k=\alpha$, will be transformed to 0 (false) or 1 (true) to represent the final decision. Finally, the maximum count of true decisions among different branches during a certain time will be output.

The structure of the DNN has also been shown in Fig. \ref{fig:subblock}, which are composed of an input layer, an output layer and three hidden layers. The activation function for the hidden layers is rectified linear unit (reLu). Furthermore, to limit the output to $[0,1]$, a sigmoid function is employed after the output layer. The detailed configuration of the DNN is summarized in Table \ref{tab:parameters}.

\begin{table}[h]
	\processtable{Configuration of the DNN in the proposed packing ratio estimation}
	{\begin{tabular}{|p{30mm}<{\centering}|p{40mm}<{\centering}|}\hline
			Item & Value \\\hline
			Node number of each layer & (20, 1000, 500, 250, 1) \\\hline
			Training data size & $3.2\times10^6$ groups \\\hline
			Training epoch & 50 \\\hline
			Optimizer & Adam \\\hline	
			Loss function & Mean square error (MSE) \\\hline
			Learning rate & 0.001 \\\hline
			Testing data size & $3.2\times10^6$ groups \\\hline
	\end{tabular}}{\label{tab:parameters}}
\end{table}

Similar to most DL methods, the application of the DNN in our proposed packing ratio estimation includes two stages named offline training and online employment. In the offline training stage, the data set is consist of many groups of 20 symbols which start with the optimal sampling points of each original symbol and are downsampled by interval $\alpha_k T_N$ from the output symbol of the matched filter. The data set is obtained from the received symbols generated with different $\alpha$ values. The label set is composed of the corresponding 0 ($\alpha\ne \alpha_k$) and 1 ($\alpha= \alpha_k$). DNN will try to learn the relevance between the input and the output and updates its parameters during each backpropagation (BP) process. Then, in the online employment stage, the DNN can independently output the probability of $\alpha=\alpha_k$ for each group of the input symbols with the well-trained parameters. Both the training and testing data are generated by software simulation.

\section{Numerical Results}
In this section, we assess the performance and robustness of our proposed DL-based packing ratio estimation for FTN signaling. Without loss of generality, some common packing ratio values are taken into consideration. The roll-off factor of the SRRC filter is 0.3. And the $E_b/N_0$ value of the training data is 4dB. Table \ref{tab:acc_of_dnn} summaries the probability of true decisions in the analyses for different $\alpha$ and possible $\alpha_k$ values. As shown, for each $\alpha$, the analysis for $\alpha_k=\alpha$ always achieve the largest number of true decisions. So, when large enough decisions have been carried out, $\alpha_k$ with the largest number of true decisions can be regarded as the accurate estimation result.

\begin{table}[h]
	\processtable{Probability of true decisions for different $\alpha$ and $\alpha_k$ values at $E_b/N_0=4dB$}
	{\begin{tabular}{|c|c|c|c|c|c|c|}\hline
			\diagbox{$\alpha$}{$P_{true}$}{$\alpha_k$} & 1 & 0.9 & 0.8 & 0.75 & 0.7 & 0.6 \\\hline
			1 & 0.7538 & 0.2181 & 0.1690 & 0.1138 & 0.1066 & 0.0952 \\\hline
			0.9 & 0.1243 & 0.7314 & 0.1598 & 0.1210 & 0.1143 & 0.0985 \\\hline
			0.8 & 0.0686 & 0.1356 & 0.6377 & 0.1416 & 0.1720 & 0.1373 \\\hline
			0.75 & 0.0451 & 0.0938 & 0.1603 & 0.5478 & 0.1970 & 0.1921 \\\hline
			0.7 & 0.0277 & 0.0681 & 0.1370 & 0.1641 & 0.5534 & 0.2797 \\\hline
			0.6 & 0.0202 & 0.0412 & 0.0953 & 0.1275 & 0.1964 & 0.5461 \\\hline
	\end{tabular}}{\label{tab:acc_of_dnn}}
\end{table}

To better illustrate the performance of the proposed scheme, we define the accuracy of the final packing ratio estimation as

\begin{align}
	P_{acc} = \sum_{m=1}^{M} \sum_{n=0}^{m-1} \boldsymbol C_M^m \boldsymbol C_M^n (p_1)^m(1-p_1)^{(M-m)}  (p_2)^n(1-p_2)^{(M-n)},
\end{align}
where $M$ is the number of decisions used to select the maximum one in Fig. \ref{fig:subblock}. $p_1$ is the probability of true decision in the analysis for $\alpha_k=\alpha$ (i.e. the diagonal of Table \ref{tab:acc_of_dnn}) while $p_2$ is the biggest one among that of analysis for $\alpha_k \ne \alpha$. Last but not least, the results in Table \ref{tab:acc_of_dnn} are obtained by simulations on the test data set.

\begin{figure}[h]
	\centering{\includegraphics[width=56mm]{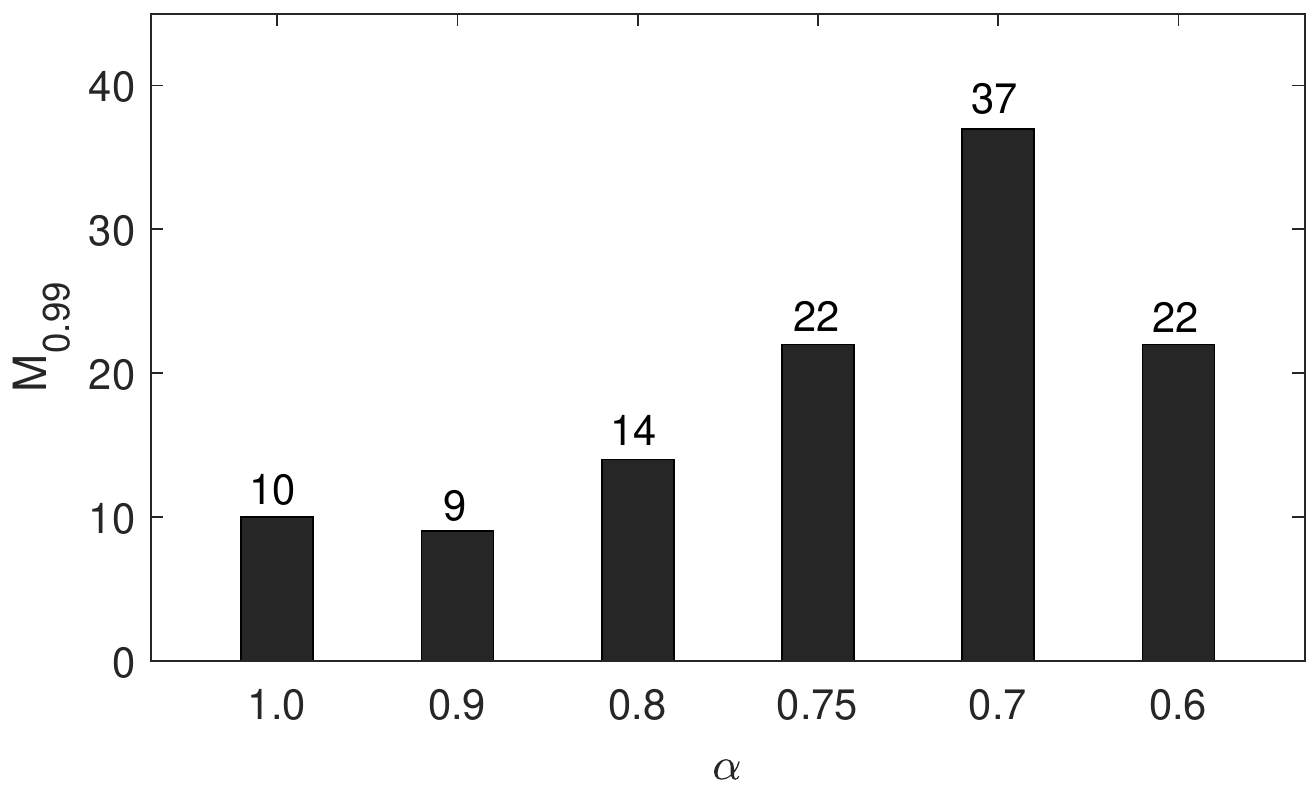}}
	\caption{The minimum number of the decisions required by the proposed estimation to achieve a 99\% accuracy at $E_b/N_0=4dB$
		\source{}}
	\label{fig:zhuzhuangtu}
\end{figure}

Here, we use $M_{0.99}$ to represents the minimum number of decisions required by the proposed packing ratio estimation for different $\alpha$ values to achieve an accuracy beyond $99\%$, which has been demonstrated in  Fig. \ref{fig:zhuzhuangtu}. As shown, the accuracy of the estimation for different $\alpha$ values can converge within less than 40 or even 10 decisions. In, practical, to guarantee the accuracy of the estimation, the biggest $M_{0.99}$ value should be employed. For example, when the packing ratio set of an adaptive FTN system is \{1, 0.9, 0.8, 0.75, 0.6\}, no less than 22 decisions must be carried out for the estimation to achieve a 99\% accuracy.

It is very important for the proposed blind estimation to be robust to the signal-to-noise ratio (SNR). Otherwise, DNNs of the proposed estimation will be trained and employed for different SNRs independently and will occupy a large complexity resulting from SNR estimation and a great number of DL network parameters that are stored for different SNRs. Fig. \ref{fig:zhuzhuangtu} illustrates the convergence speed of the proposed estimation for $\alpha=0.8$ which are trained at $E_b/N_0=4dB$ and employed for different SNRs. As shown, after sufficient training, the proposed estimation can perform effective convergences in the accuracy at various $E_b/N_0$ values.

\begin{figure}[h]
	\centering{\includegraphics[width=56mm]{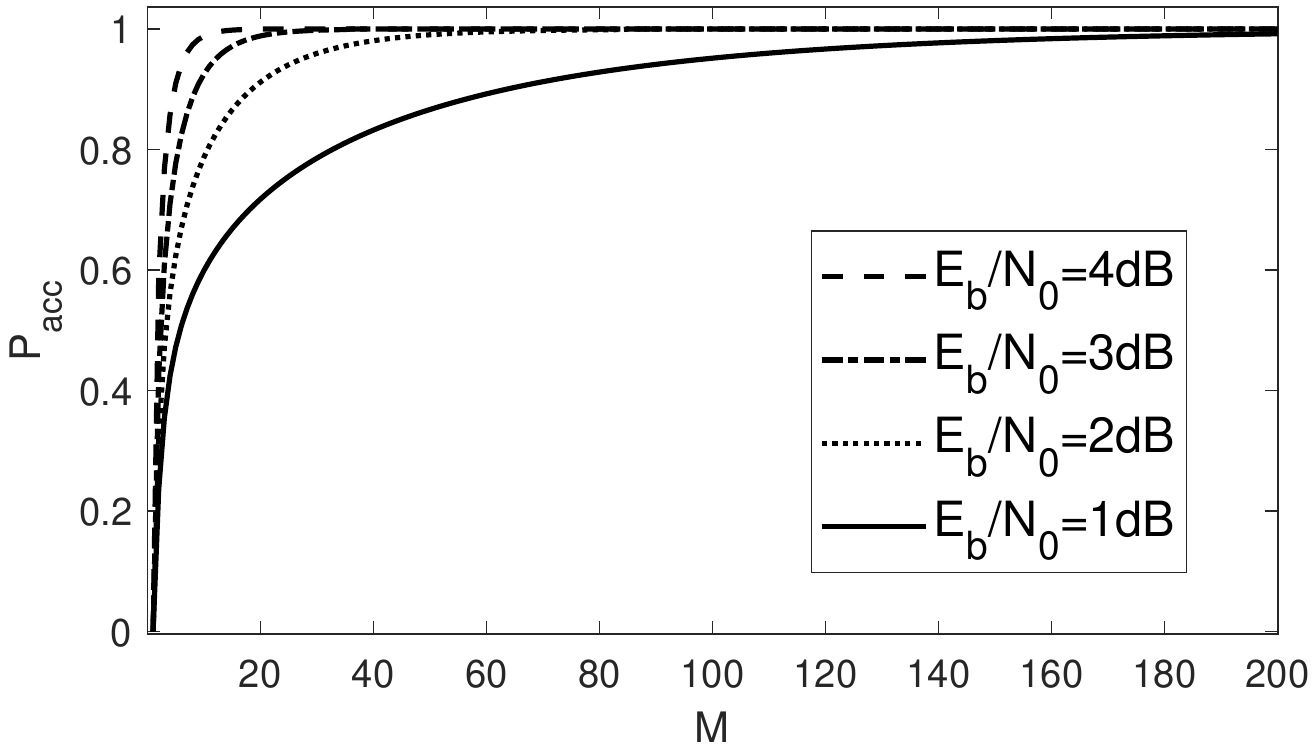}}
	\caption{The convergence speed in the accuracy of the proposed blind  estimation for $\alpha=0.8$ which are trained at $E_b/N_0=4dB$ and employed for other $E_b/N_0$ values
		\source{}}
	\label{fig:convergence}
\end{figure}

\section{Summary and Conclusion}
In this work, a DL-based blind packing ratio estimation for FTN signaling is proposed. Simulation results have shown its fast convergence in accuracy and the robustness to SNR of the proposed estimation. For packing-ratio-based adaptive FTN signaling, the proposed estimation can replace the dedicated channel to further improve the spectrum utilization. Also, for the FTN-aided communications with unknown packing ratio, the proposed estimation will help the eavesdropper get the real packing ratio from the received symbols and then recover the origin signals.


\end{document}